# Nature of Valance Band Splitting on Multilayer MoS$_2$

Xiaofeng Fan$^{a,}$ *, W.T. Zheng$^{a}$, and David J. Singh$^{a,b,}$ †

a. College of Materials Science and Engineering, Jilin University, Changchun 130012, China

b. Department of Physics and Astronomy, University of Missouri, Columbia, Missouri 65211-7010, USA

*E-mail: xffan@jlu.edu.cn; † E-mail: singhdj@missouri.edu

## Abstract

Understanding the origin of splitting of valance band is important since it governs the unique spin and valley physics in few-layer MoS$_2$. With first principle methods, we explore the effects of spin-orbit coupling and layer's coupling on few-layer MoS$_2$. It is found that intra-layer spin-orbit coupling has a major contribution to the splitting of valance band at K. In double-layer MoS$_2$, the layer's coupling results in the widening of energy gap of splitted states induced by intra-layer spin-orbit coupling. The valance band splitting of bulk MoS$_2$ in K can follow this model. We also find the effect of inter-layer spin-orbit coupling in triple-layer MoS$_2$. In addition, the inter-layer spin-orbit coupling is found to become to be stronger under the pressure and results in the decrease of main energy gap in the splitting valance bands at K. .

## Introduction

A new class of 2D materials, the single-layer and/or few-layer of hexagonal transition metal dichalcogenides (h-TMDs), have attracted broad attentions due to the extraordinary physical properties and promising applications in electric and optoelectronic devices[1-5]. As the prototypical 2D materials, single-layer h-TMDs are direct band gap semiconductors with spin-splitting at valance band maximum which is much different from graphene[6-9]. This promises a chance to manipulate the spin degree of freedom and valley polarization[10-12]. In addition, with extreme dimensional confinement, tightly-bound excitons and strong electron-electron interactions due to weak screening, h-TMDs have been ideal low-dimensional compounds to explore many interesting quantum phenomena[11, 13-16], such as spin- and valley- Hall effects and superconductivity[17-19]. There are also a lot of fascinating optical properties in single-layer h-TMDs, such as the strong band gap photoluminescence at edge[5], surface sensitive luminescence[20, 21] and strain-controlled optical band gap[22-25], and so on.

Among these h-TMDs, MoS$_2$ is a representative. Bulk MoS$_2$ is a layered compound stacked with the weak van der Waals interaction[26]. Due to the highly anisotropic mechanical property, it is used in dry lubrication. It has also made the interest due to the special catalytic activity from its edge[27]. In each layer of MoS$_2$, there are three atomic layers with a center layer of Mo around S layers in both sides. The states near band gap are well-known to be mainly from the d-orbitals of Mo[28]. There is a priori proposal that the layer's coupling is possible to have a very weak effect to the states near band gap. Bulk $MoS_2$ is an indirect-gap semiconductor with a band gap of 1.29 eV. However, following the reduction of layers to sinlge-layer, there is a transition between indirect band gap and direct gap[3]. The single-layer $MoS_2$ is found to have a direct band gap of about 1.8 eV[5, 29]. Therefore, the layer's coupling has a strong effect to the states near band gap with recent reports[30]. Especially, the states of valance band top at Γ point (VB-Γ) and conduction band bottom along Λ (CB-Λ) is much sensitive to the layer's coupling (LC). Compared with the states of VB-Γ and CB- Λ, the LC effects on the states of valance band top and conduction band bottom at K point (VB-K, CB-K) are very weak. Therefore, there remains an open question about the origin of the splitting at valance band of K point which governs the unique spin and valley physics. In the single-layer limit, the splitting can be attributed entirely to spin-orbit coupling (SOC). In bulk limit, it is considered to be a result of combination of SOC and LC. However, there is disagreement about the relative strength of both mechanisms[31-37].

In the work, we explore the effect of SOC on few-layer $MoS_2$ with the rule of LC by first principle methods in details. We analyze the splitting of states at VB-Γ, VB-K CB-Λ, and CB-K and explore the change of splitting by following the increase of distance of both layers for double-layer $MoS_2$. It is found that intra-layer SOC (intra-SOC) has a major contribution to the splitting at VB-K, while LC can open effectively the degeneracy of states at VB-K. With the analysis of charge distribution in real space, the double-degeneracy of states at the valance band maximum of K point, which isn't broken due to the inter-layer inverse symmetry for both layers which results in the forbidding of inter-layer SOC, are mainly from the spin-up state of first-layer and spin-down state of second layer. For triple-layer $MoS_2$, the LC with inter-layer SOC due to the absence of inter-layer inverse symmetry in three-layer system makes the splitting complicated. The intra-layer SOC results in two main bands splitting, while in each main band, the triple-degeneracy is broken mainly due to the inter-layer SOC. With the pressure, it is found in double-layer $MoS_2$ that the double-degeneracy of states in each main band isn't broken when the splitting of both main bands is increased due to the strengthening of LC. For triple-layer $MoS_2$ under large pressure, the splitting of triple-degeneracy in each main band is very obvious.

**Computational Method**

The present calculations are performed within density functional theory using accurate frozen-core full-potential projector augmented-wave (PAW) pseudopotentials, as implemented in the VASP code[38-40]. The generalized gradient approximation (GGA) with the parametrization of Perdew-Burke-Ernzerhof (PBE) and with added van der Waals corrections is used[41]. The *k*-space integrals and the plane-wave basis sets are chosen to ensure that the total energy is converged at the 1 meV/atom level. A kinetic energy cutoff of 500 eV for the plane wave expansion is found to be sufficient. The effect of dispersion interaction is included by the empirical correction scheme of Grimme (DFT+D/PBE)[42]. This approach has been successful in describing layered structures[43, 44].

The lattice constants *a* and *c* of bulk $MoS_2$ are 3.191 Å and 12.374 Å which is similar to that from the experiments (3.160 Å and 12.295 Å). For the different layered MoS$_2$, the supercells are constructed with a vacuum space of 20 Å along z direction. The Brillouin zones are sampled with the Γ-centered k-point grid of 18×18×1. With the state-of-the-art method of adding the stress to stress tensor in VASP code[39, 40], the structure of bulk MoS$_2$ is optimized under a specified hydrostatic pressure of 15GPa. With these structural parameters from bulk MoS$_2$, the double- and triple-layer MoS$_2$ structures under the pressures are constructed. The electronic properties can analyzed with and/or without spin-orbit coupling to explore the band splitting near band gap. The calculated band gap of single-layer MoS$_2$ without the consideration of spin-orbit interaction is 1.66 eV and less than the experimental report of about 1.8 eV. Obviously, the band gap from PBE is underestimated as in common in usual density functional calculations. Though the band gap is underestimated by PBE, the band structure near Fermi level doesn't have obvious difference from that from other many body methods.

**Results and discussion**

The structure of single-layer MoS$_2$ has the hexagonal symmetry with space group *P-6m2*. The six sulfur atoms near each Mo atom form a trigonal prismatic structure with the mirror symmetry in *c* direction. Obviously, the reversal symmetry is absent, and the intra-SOC in the band structure becomes to be free. An obvious band splitting on the valance band maximum around K (K') point has been observed and is contributed to the SOC. In addition, the SOC also results in the band splitting on conduction band minimum around Λ point, while the splitting at VB-Γ and CB-K is not opened. The states of VB-Γ and CB-K are contributed mostly from $d_z^2$ orbital of Mo and the effect of the spin-orbit effect is very weak. At the same time, the states of VB-K and CB-Λ are mainly from $d_{x^2-y^2}$ and $d_{xy}$ orbitals of Mo and the spin-orbit effect on Mo can be revealed in the case of the absence of reversal symmetry. The band splitting at VB-K (149 meV) is larger than that at CB-Λ (about 79 meV). It may be that the distribution of change (or wave function) at CB-Λ around Mo atoms in xy plane is more localized than that at VB-K[30]. We also notice that the charge distribution of spin-up state is much different from that of spin-down state at VB-K. The spin-down state around Mo is more localized than the spin-up state.

For double-layer MoS$_2$, the interaction between two layers becomes important to the states near Fermi level. One of much evident effects is the direct band gap (K-K) of single-layer becomes to be indirect band gap (Γ-K) due to the uplift of state at VB-Γ. It can be ascribed to large band splitting (0.618 eV) at VB-Γ. It is also observed that the band splitting at the conduction band bottom around Λ is about 0.352 eV. However, without the consideration of SOC effect, the band splitting at VB-K is just 73.8 meV. The difference of LC's strength of different states at VB-Γ, VB-K and CB-Λ is ascribed to the charge distribution near sulfur atoms. The large contribution of charge on sulfur atoms for the states at VB-Γ makes the LC to become easier[30]. The weak LC at VB-K may make the SOC important. In order to explore the rules of SOC and LC in double-layer MoS$_2$, we calculated the change of band structures by following the change of distance between both layers with and without the consideration of spin-orbit effect. As shown in Fig. 1, the band splitting including that at VB-Γ, VB-K and CB-Λ approaches zero quickly following the increase of distance, especially that of VB-K, if the spin-orbit effect is not considered. With SOC, the band splitting at VB-K and CB-Λ converge towards some constants (about 149 meV and 79 meV), while the band splitting at VB-Γ approaches zero. Obviously, there is no spin-orbit effect at VB-Γ. With the large distance between both layers, the effect of LC can be ignored and the splitting is from SOC.

It has been well-known that the spin-up and spin-down states at VB-K' are reversed by compared with that at VB-K in single-layer $MoS_2$. For double-layer $MoS_2$, both splitting bands at VB-K are two-degeneracy. As shown in Fig. 2c, the upper band of both bands is composed by the spin-up state of first layer and spin-down state of second layer (~|1↑⟩, ~|2↓⟩). The lower band is with the spin-down state of first layer and spin-up state of second layer (~|1↓⟩, ~|2↑⟩). Obviously, the energies of spin-up and spin-down of second layer at VB-K are reversed by compared with that of first layer. Since there is reversal symmetry for double-layer system, the energies of states |1↑⟩ and |2↓⟩ with same energy cannot be split due to the absence of inter-layer SOC (inter-SOC). Therefore, we can understand the splitting at VB-K based on the intra-SOC and LC with the theoretical model shown in Fig. 2a. Because of the splitting of intra-SOC, the energies of |1↑⟩ and |2↑⟩ are very different. This reduces largely the coupling of both states due to layer's interaction. From the band splitting value (166 meV) of double layer at VB-K, the increased splitting from LC effect is about 17 meV and is much less than that (73.8 meV) from LC without the consideration of spin-orbit effect. For the spin-down channel, the mechanism of band splitting is same to that of spin-up channel. Therefore, the contribution of intra-SOC (149 meV) to the band splitting at VB-K is much larger than that of LC. The same mechanism about the splitting at VB-K (shown in Fig. 2a) can be used for bulk $MoS_2$. The contribution of LC is increased to about 59 meV since the band splitting at VB-K is about 0.208 eV. If no considering the spin-orbit effect, the band splitting due to LC is about 145.7 meV which much similar to the value from SOC(149 meV) in single-layer. This may be the reason that there is disagreement about the relative strength of both effects in bulk limit. Based on the model mentioned above and analysis, the intra-SOC effect is the main mechanism for the splitting at VB-K in bulk limit.

For triple-layer $MoS_2$, the band splitting near band gap is complicated, since there are three states from three layers which are coupling with each other and hybridized with possible inter-SOC. For the states at VB-Γ, there is no SOC effect and the three degenerate states will be splitting due to LC. It is found that the two splitting values ($\Delta_\Gamma^1$ and $\Delta_\Gamma^2$ in Fig. 3a) which control the relative energy difference of three states after the hybridization are 0.293 eV and 0.502 eV, respectively. The much different value of both splitting implies that there is strong coupling between first layer and third layer since both splitting values should be equivalent if the nearest-neighbor interaction is just considered for the three degenerate states. Without the consideration of spin-orbit effect, the splitting values at CB-Λ ($\Delta_\Lambda^1$ and $\Delta_\Lambda^2$) are 0.241 eV and 0.225 eV and that at VB-K ($\Delta_K^1$ and $\Delta_K^2$) are 49 meV and 55 meV, respectively. Based on the nearest-neighbor LC strength (73.8/2 meV) at VB-K from double layer, the LC strength between first-layer and third layer of triple layer is about 2 meV at VB-K and may be ignored. Therefore, we propose a coupling model based on the intra-SOC and nearest-neighbor LC, as shown in Fig. 3c. With this model, the spin-up and spin-down bands of each layer are splitted by the intra-SOC. Then the LC will perturb these states for each spin channel. For example, the spin-up states are composed with two degenerate upper states ($|1\uparrow\rangle$, $|3\uparrow\rangle$) and one lower state ($|2\uparrow\rangle$) in Fig. 3c and LC will result in the splitting of two degenerate upper states with the increase of energy gap between $|2\uparrow\rangle$ and $|3\uparrow\rangle$. With the spin-up and spin-down channels together after LC, there should two main bands and each main band is composed with two degenerate states and one single states. The LC doesn't change the energy gap between the main bands. It is found that the energy gap $\Delta'_{SOC}$ is 148.7 meV and similar to the splitting from intra-SOC ($\Delta_{SOC}$ =149 meV). However, it is interesting that the two degenerate states in each main band, such as the upper states $\sim|1\uparrow\rangle$ and $\sim|2\downarrow\rangle$ and the lower states $\sim|2\uparrow\rangle$ and $\sim|3\downarrow\rangle$, are splitted, as shown in the inset of Fig. 3c. In addition, it is found that the splitting values are so large (such as, 11.3 meV between $\sim|1\uparrow\rangle$ and $\sim|2\downarrow\rangle$) that the contribution of LC between first layer and third layer is not enough. We propose that the splitting of degenerate states in each main band is from the inter-SOC.

While the small pressure doesn't induce the obvious splitting from intra-SOC in single-layer $MoS_2$, it is possible there is strong effect to the splitting from inter-SOC in triple-layer $MoS_2$. For double-layer $MoS_2$, the inter-SOC is forbidden and the degenerate state in each main band isn't opened and the energy gap between both main bands is increased with the strengthening of LC under the pressure in Fig. 4a. In triple-layer, the strengthening of LC under pressure should have the obvious effect, such as the increase of energy gap between $\sim|1\uparrow\rangle$ and $\sim|3\uparrow\rangle$ in Fig. 4c. Besides the enhanced LC effect, an apparent observation is the energy gap between main bands $\Delta'_{SOC}$ has been decreased to 135.8 meV under 15 GPa in Fig. 4b. This should be the typical evidence for the inter-SOC.

**Conclusions**

We study the band splitting at valance band maximum of multi-layer $MoS_2$ by first principle methods in details. We propose a model based on the intra-layer spin-orbit coupling to solve the valance band splitting at K point of multi-layer $MoS_2$ and bulk $MoS_2$ with the perturbation of layer's coupling and inter-layer spin-orbit coupling. It is also found that the direct interaction between second near-neighbor layers is weak at VB-K. While the inter-layer spin-orbit coupling is forbidden in double-layer $MoS_2$, this effect appears in triple-layer $MoS_2$. Especially, under the pressure, the inter-layer spin-orbit coupling is raised with the decrease of energy gap between main bands from intra-layer spin-orbital coupling.

**References**

1. Wang, Q. H.; Kalantar-Zadeh, K.; Kis, A.; Coleman, J. N.; Strano, M. S., Electronics and Optoelectronics of Two-Dimensional Transition Metal Dichalcogenides. *Nat. Nanotech.* **2012,** 7, 699-712.
2. Radisavljevic, B.; Radenovic, A.; Brivio, J.; Giacometti, V.; Kis, A., Single-Layer $MoS_2$ Transistors. *Nat Nanotech.* **2011,** 6, 147-150.
3. Mak, K. F.; Lee, C.; Hone, J.; Shan, J.; Heinz, T. F., Atomically Thin $MoS_2$: A New Direct-Gap Semiconductor. *Phys. Rev. Lett.* **2010,** 105, 136805.
4. Lee, C.; Yan, H.; Brus, L. E.; Heinz, T. F.; Hone, J.; Ryu, S., Anomalous Lattice Vibrations of Single- and Few-Layer $MoS_2$. *ACS Nano* **2010,** 4, 2695-2700.

5. Splendiani, A.; Sun, L.; Zhang, Y.; Li, T.; Kim, J.; Chim, C.-Y.; Galli, G.; Wang, F., Emerging Photoluminescence in Monolayer $MoS_2$. *Nano Lett.* **2010,** 10, 1271-1275.

6. Zhang, Y.; Chang, T.-R.; Zhou, B.; Cui, Y.-T.; Yan, H.; Liu, Z.; Schmitt, F.; Lee, J.; Moore, R.; Chen, Y.; Lin, H.; Jeng, H.-T.; Mo, S.-K.; Hussain, Z.; Bansil, A.; Shen, Z.-X., Direct observation of the transition from indirect to direct bandgap in atomically thin epitaxial MoSe2. *Nat Nanotech* **2014,** 9, 111-115.

7. Yeh, P.-C.; Jin, W.; Zaki, N.; Zhang, D.; Liou, J. T.; Sadowski, J. T.; Al-Mahboob, A.; Dadap, J. I.; Herman, I. P.; Sutter, P.; Osgood, R. M., Layer-dependent electronic structure of an atomically heavy two-dimensional dichalcogenide. *Phys. Rev. B* **2015,** 91, 041407.

8. Zhu, Z. Y.; Cheng, Y. C.; Schwingenschl枚gl, U., Giant spin-orbit-induced spin splitting in two-dimensional transition-metal dichalcogenide semiconductors. *Phys. Rev. B* **2011,** 84, 153402.

9. Zeng, H.; Dai, J.; Yao, W.; Xiao, D.; Cui, X., Valley polarization in $MoS_2$ monolayers by optical pumping. *Nat. Nanotech.* **2012,** 7, 490-493.

10. Cao, T.; Wang, G.; Han, W.; Ye, H.; Zhu, C.; Shi, J.; Niu, Q.; Tan, P.; Wang, E.; Liu, B.; Feng, J., Valley-selective circular dichroism of monolayer molybdenum disulphide. *Nat Commun* **2012,** 3, 887.

11. Xiao, D.; Liu, G.-B.; Feng, W.; Xu, X.; Yao, W., Coupled Spin and Valley Physics in Monolayers of MoS2 and Other Group-VI Dichalcogenides. *Phys. Rev. Lett.* **2012,** 108, 196802.

12. Yao, W.; Xiao, D.; Niu, Q., Valley-Dependent Optoelectronics from Inversion Symmetry Breaking. *Phys. Rev. B* **2008,** 77, 235406.

13. Klots, A. R.; Newaz, A. K. M.; Wang, B.; Prasai, D.; Krzyzanowska, H.; Lin, J.; Caudel, D.; Ghimire, N. J.; Yan, J.; Ivanov, B. L.; Velizhanin, K. A.; Burger, A.; Mandrus, D. G.; Tolk, N. H.; Pantelides, S. T.; Bolotin, K. I., Probing excitonic states in suspended two-dimensional semiconductors by photocurrent spectroscopy. *Scientific Reports* **2014,** 4, 6608.

14. Li, X.; Zhang, F.; Niu, Q., Unconventional Quantum Hall Effect and Tunable Spin Hall Effect in Dirac Materials: Application to an Isolated $MoS_2$ Trilayer. *Phys. Rev. Lett.* **2013,** 110, 066803.

15. Ross, J. S.; Wu, S.; Yu, H.; Ghimire, N. J.; Jones, A. M.; Aivazian, G.; Yan, J.; Mandrus, D. G.; Xiao, D.; Yao, W.; Xu, X., Electrical control of neutral and charged excitons in a monolayer semiconductor. *Nat Commun* **2013,** 4, 1474.

16. Mak, K. F.; He, K.; Lee, C.; Lee, G. H.; Hone, J.; Heinz, T. F.; Shan, J., Tightly bound trions in monolayer $MoS_2$. *Nat Mater* **2013,** 12, 207-211.

17. Mak, K. F.; McGill, K. L.; Park, J.; McEuen, P. L., The valley Hall effect in MoS2 transistors. *Science* **2014,** 344, 1489-1492.

18. Roldán, R.; Cappelluti, E.; Guinea, F., Interactions and superconductivity in heavily doped MoS${}_{2}$. *Phys. Rev. B* **2013,** 88, 054515.

19. Geim, A. K.; Grigorieva, I. V., Van der Waals heterostructures. *Nature* **2013,** 499, 419-425.

20. Mouri, S.; Miyauchi, Y.; Matsuda, K., Tunable Photoluminescence of Monolayer $MoS_2$ via Chemical Doping. *Nano Letters* **2013**, (12), 5944-5948.

21. Tongay, S.; Suh, J.; Ataca, C.; Fan, W.; Luce, A.; Kang, J. S.; Liu, J.; Ko, C.; Raghunathanan, R.; Zhou, J.; Ogletree, F.; Li, J.; Grossman, J. C.; Wu, J., Defects activated photoluminescence in two-dimensional semiconductors: interplay between bound, charged, and free excitons. *Scientific Reports* **2013,** 3, 2657.

22. Feng, J.; Qian, X.; Huang, C.-W.; Li, J., Strain-Engineered Artificial Atom as a Broad-Spectrum Solar Energy Funnel. *Nat. Photon.* **2012,** 6, 866-872.

23. Fan, X.; Zheng, W.; Kuo, J.-L.; Singh, D. J., Structural Stability of Single-layer $MoS_2$ under Large Strain. *J. Phys.: Condens. Matter* **2015,** 27, 105401.

24. Scalise, E.; Houssa, M.; Pourtois, G.; Afanasev, V.; Stesmans, A., Strain-Induced Semiconductor to

Metal Transition in the Two-Dimensional Honeycomb Structure of MoS$_2$. *Nano Research* **2012,** 5, 43-48.

25. Conley, H. J.; Wang, B.; Ziegler, J. I.; Haglund, R. F.; Pantelides, S. T.; Bolotin, K. I., Bandgap Engineering of Strained Monolayer and Bilayer MoS2. *Nano Lett.* **2013,** 13, 3626-3630.

26. Mattheiss, L. F., Band Structures of Transition-Metal-Dichalcogenide Layer Compounds. *Phys. Rev. B* **1973,** 8, 3719-3740.

27. Kibsgaard, J.; Chen, Z.; Reinecke, B. N.; Jaramillo, T. F., Engineering the Surface Structure of MoS$_2$ to Preferentially Expose Active Edge Sites for Electrocatalysis. *Nat. Mater.* **2012,** 11, 963-969.

28. Chang, C.-H.; Fan, X.; Lin, S.-H.; Kuo, J.-L., Orbital analysis of electronic structure and phonon dispersion in MoS$_2$, MoSe$_2$, WS$_2$, and WSe$_2$ monolayers under strain. *Phys. Rev. B* **2013,** 88, 195420.

29. Kuc, A.; Zibouche, N.; Heine, T., Influence of Quantum Confinement on the Electronic Structure of the Transition Metal Sulfide TS$_2$. *Phys. Rev. B* **2011,** 83, 245213.

30. Fan, X.; Chang, C. H.; Zheng, W. T.; Kuo, J.-L.; Singh, D. J., The Electronic Properties of Single-Layer and Multilayer MoS2 under High Pressure. *J. Phys. Chem. C* **2015,** 119, 10189-10196.

31. Latzke, D. W.; Zhang, W.; Suslu, A.; Chang, T.-R.; Lin, H.; Jeng, H.-T.; Tongay, S.; Wu, J.; Bansil, A.; Lanzara, A., Electronic structure, spin-orbit coupling, and interlayer interaction in bulk MoS$_2$ and WS$_2$. *Phys. Rev. B* **2015,** 91, 235202.

32. Klein, A.; Tiefenbacher, S.; Eyert, V.; Pettenkofer, C.; Jaegermann, W., Electronic band structure of single-crystal and single-layer WS$_2$: Influence of interlayer van der Waals interactions. *Phys. Rev. B* **2001,** 64, 205416.

33. Molina-Sanchez, A.; Sangalli, D.; Hummer, K.; Marini, A.; Wirtz, L., Effect of Spin-Orbit Interaction on the Optical Spectra of Single-Layer, Double-Layer, and Bulk MoS$_2$. *Phys. Rev. B* **2013,** 88, 045412.

34. Alidoust, N.; Bian, G.; Xu, S.-Y.; Sankar, R.; Neupane, M.; Liu, C.; Belopolski, I.; Qu, D.-X.; Denlinger, J. D.; Chou, F.-C.; Hasan, M. Z., Observation of monolayer valence band spin-orbit effect and induced quantum well states in MoX$_2$. *Nat Commun* **2014,** 5, 1312.

35. Eknapakul, T.; King, P. D. C.; Asakawa, M.; Buaphet, P.; He, R. H.; Mo, S. K.; Takagi, H.; Shen, K. M.; Baumberger, F.; Sasagawa, T.; Jungthawan, S.; Meevasana, W., Electronic Structure of a Quasi-Freestanding MoS$_2$ Monolayer. *Nano Letters* **2014,** 14, 1312-1316.

36. Jin, W.; Yeh, P.-C.; Zaki, N.; Zhang, D.; Sadowski, J. T.; Al-Mahboob, A.; van der Zande, A. M.; Chenet, D. A.; Dadap, J. I.; Herman, I. P.; Sutter, P.; Hone, J.; Osgood, R. M., Direct Measurement of the Thickness-Dependent Electronic Band Structure of MoS$_2$ Using Angle-Resolved Photoemission Spectroscopy. *Phys. Rev. Lett.* **2013,** 111, 106801.

37. Suzuki, R.; Sakano, M.; Zhang, Y. J.; Akashi, R.; Morikawa, D.; Harasawa, A.; Yaji, K.; Kuroda, K.; Miyamoto, K.; Okuda, T.; Ishizaka, K.; Arita, R.; Iwasa, Y., Valley-dependent spin polarization in bulk MoS$_2$ with broken inversion symmetry. *Nat Nano* **2014,** 9, 611-617.

38. Hohenberg, P.; Kohn, W., Inhomogeneous Electron Gas. *Phys. Rev.* **1964,** 136, B864.

39. Kresse, G.; Furthmüller, J., Efficient Iterative Schemes for Ab Initio Total-Energy Calculations Using a Plane-wave Basis Set. *Phys. Rev. B* **1996,** 54, 11169–11186

40. Kresse, G.; Furthmüller, J., Efficiency of Ab-Initio Total Energy Calculations for Metals and Semiconductors Using a Plane-wave Basis Set *Computat. Mater. Sci.* **1996,** 6, 15-50.

41. Perdew, J. P.; Burke, K.; Ernzerhof, M., Generalized Gradient Approximation Made Simple. *Phys. Rev. Lett.* **1996,** 77, 3865-3868.

42. Grimme, S., Semiempirical GGA-Type Density Functional Constructed with a Long-Range Dispersion Correction. *J. Comput. Chem.* **2006,** 27, 1787.

43. Fan, X. F.; Zheng, W. T.; Chihaia, V.; Shen, Z. X.; Kuo, J.-L., Interaction Between Graphene and the

Surface of SiO$_2$. *J. Phys.: Condens. Matter* **2012,** 24, 305004.

44. Mercurio, G.; McNellis, E. R.; Martin, I.; Hagen, S.; Leyssner, F.; Soubatch, S.; Meyer, J.; Wolf, M.; Tegeder, P.; Tautz, F. S.; Reuter, K., Structure and Energetics of Azobenzene on Ag(111): Benchmarking Semiempirical Dispersion Correction Approaches. *Phys. Rev. Lett.* **2010,** 104, 036102.

Fig. 1.

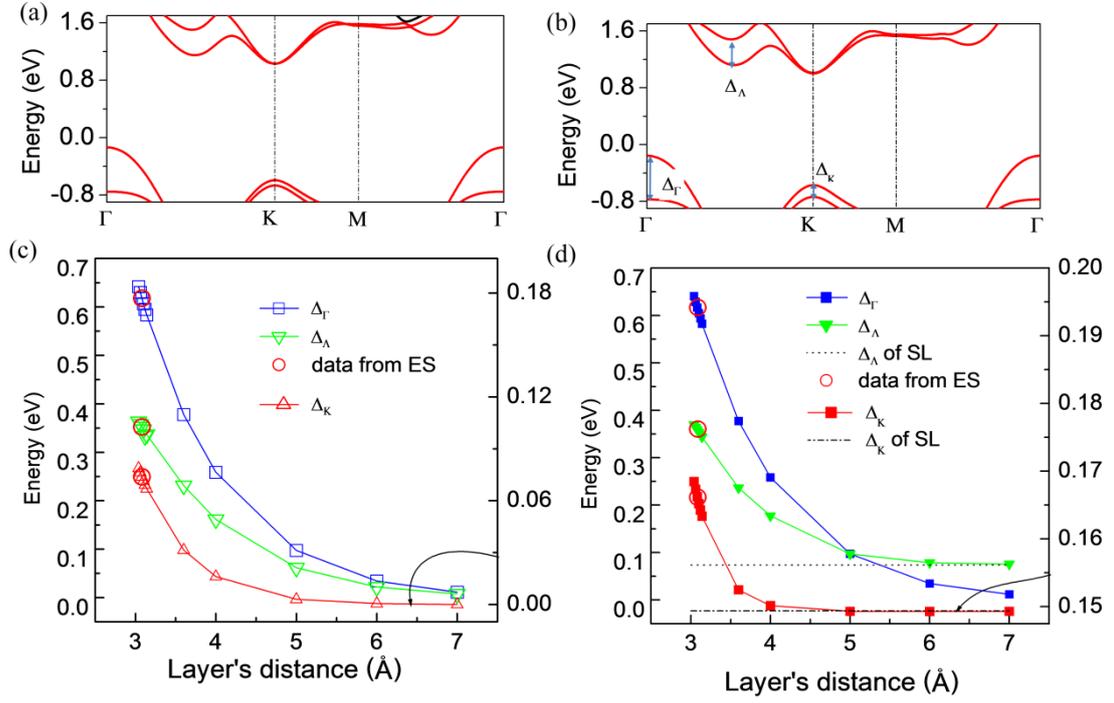

Fig.1 Band structure of double-layer MoS$_2$ calculated without spin-orbit coupling (a) and with spin-orbit coupling, and the changes of conduction band splitting ($\Delta_\Lambda$) at $\Lambda$ point and valance band splitting at $\Gamma$ point ($\Delta_\Gamma$) and K point ($\Delta_K$) following the distance between two layers of double-layer MoS$_2$, calculated without spin-orbit coupling (c) and with spin-orbit coupling (d). Note the red circles in Fig. 1c and d represent the data from the equilibrium (or stable) state (ES) and the dot and dash dot lines present the conduction band splitting ($\Delta_\Lambda$) at $\Lambda$ point and valance band splitting ($\Delta_K$) at K point of single-layer MoS$_2$, respectively.

Fig.2

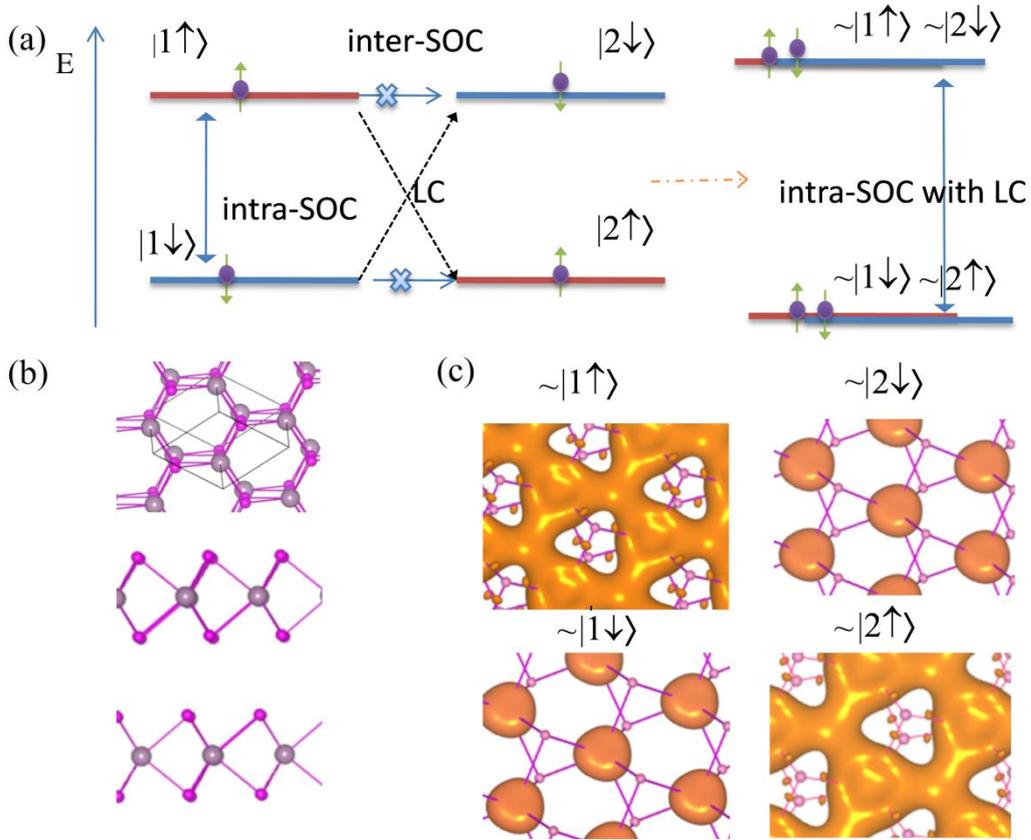

Fig. 2 Schematic of valance band splitting of valance band maximum at K point due to the spin-orbit coupling in the each layer (intra-SOC) and layer's coupling (LC) in band structure of double-layer $MoS_2$ (a), schematic structure of double-layer $MoS_2$ (b), the isosurface of band-decomposed charge density of four states at valance band maximum of K point including the states ~$|1\uparrow\rangle$, ~$|2\downarrow\rangle$, ~$|1\downarrow\rangle$ and ~$|2\uparrow\rangle$ shown in Fig. 2a after considering the effects of intra-SOC and LC.

Fig. 3

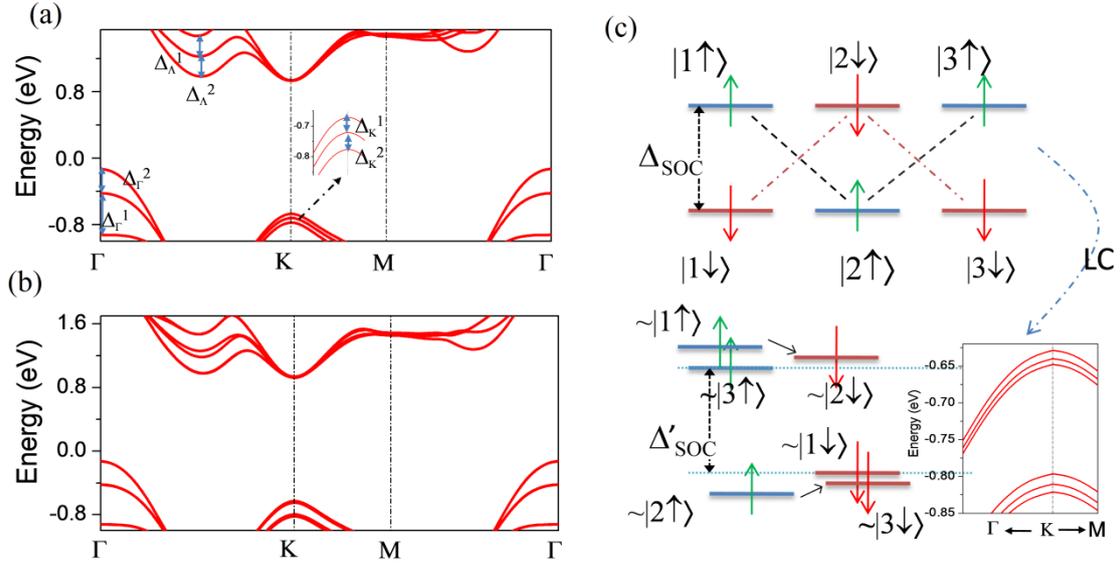

Fig. 3 Band structure of triple-layer MoS$_2$ calculated without spin-orbit coupling (a) and with spin-orbit coupling (b), Schematic of valance band splitting of valance band maximum at K point due to the spin-orbit (SOC) and layer's coupling (LC) in band structure of triple-layer MoS$_2$ (c). Note that in the inset of Fig. 3c, the band structure is plotted with two directions K→Γ and K→M and the lengths for K→Γ and K→M are the 1/10 of total lengths in the two directions, respectively.

Fig. 4

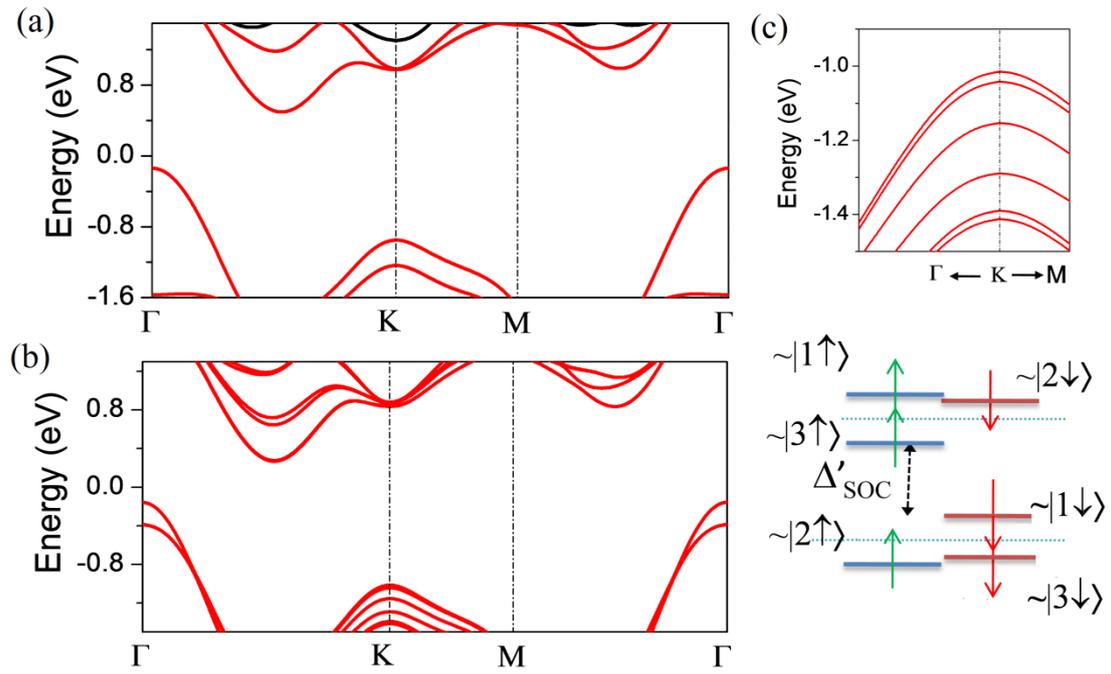

Fig. 4 Band structures of double-layer MoS$_2$ (a) and triple-layer MoS$_2$ (b) under the pressure of 15 GPa calculated with spin-orbit coupling, and valance bands of triple-layer MoS$_2$ under 15 GPa near K Point and the related schematic about band splitting due to the spin-orbit (SOC) and layer's coupling (LC).